\documentclass[twocolumn,amsmath,amssymb,prb,letterpaper,superscriptaddress,floatfix]{revtex4}
\usepackage{graphicx}
\usepackage{color}

\newcommand{\bb}{\mathbb}
 
\begin{document}

\title{Generating derivative structures: Algorithm and applications}
\begin{abstract}
  We present an algorithm for generating all derivative
  superstructures---for arbitrary parent structures and for any number
  of atom types. This algorithm enumerates superlattices and atomic
  configurations in a geometry-independent way. The key concept is to
  use the quotient group associated with each superlattice to
  determine all unique atomic configurations. The run time of the
  algorithm scales \emph{linearly} with the number of unique
  structures found. We show several applications demonstrating how the
  algorithm can be used in materials design problems. We predict an
  altogether new crystal structure in Cd-Pt and Pd-Pt, and several new
  ground states in Pd-rich and Pt-rich binary systems.
\end{abstract}

\author{Gus L. W. Hart} \affiliation{Department of Physics \& Astronomy, Brigham
  Young University, Provo Utah 84602 USA}

\author{Rodney W. Forcade} \affiliation{Department of Mathematics, Brigham
  Young University, Provo Utah 84602 USA}


\date{\today} 

\maketitle 

\section{Why derivative structures?}
Derivative superstructures\cite{buerger1947} play an important role in different
materials phenomenon such as chemical ordering in alloys, spin ordering in
magnets, and vacancy ordering in non-stoichiometric materials.  Similarly,
derivative super\emph{lattices}\cite{note2,Santoro:a09671,Santoro:a08677} are
important in problems such as twinning. What is a derivative superstructure? A
derivative superstructure is a structure whose lattice vectors are multiples of
those of a `parent lattice' and whose atomic basis vectors correspond to
\emph{lattice points} of the parent lattice. Many structures
of intermetallic compounds can be classified as fcc-derived
superstructures. These superstructures have atomic sites that closely correspond
to the sites of an fcc lattice but some of the translational symmetry is broken
by a periodic arrangement of different kinds of atoms. The structures shown in
Fig.~\ref{fig:17} comprise the set of all fcc-derived binary superstructures
with unit cell sizes 2, 3, and 4 times larger than the parent lattice.

Large sets of derivative superstructures are often used in
(practically) exhaustive searches of binary configurations on a
lattice to determine ground state properties of intermetallic
systems. The approach is not limited to searches of configurational
energies, but other physical observables can also be targeted if an
appropriate Hamiltonian is available. For example, Kim et
al.\cite{kkim:05} used an empirical pseudopotential Hamiltonian and a
large list of derivative superstructures to directly search semiconductor
alloys for desirable band-gaps and effective masses. The set of
derivative superstructures is useful in any situation where the
physical observable of interest depends on the atomic
configuration. 

For the aforementioned reasons, an algorithm for systematically generating all
superstructures of a given parent structure is useful. Such an algorithm has
been presented in the literature only once\cite{lgferreira:91} (FWZ below), but
closely related algorithms have been implemented in several alloy modeling
packages.\cite{avdwalle:02,avdwalle:02b,zarkevich:104203} The FWZ algorithm
is restricted to fcc- and bcc-based superstructures and to binary cases
only. Furthermore, the list generated by the FWZ algorithm is formally
incomplete (though in practice it may be sufficient).\cite{note3} 

\begin{figure}
\includegraphics[width=.45\linewidth]{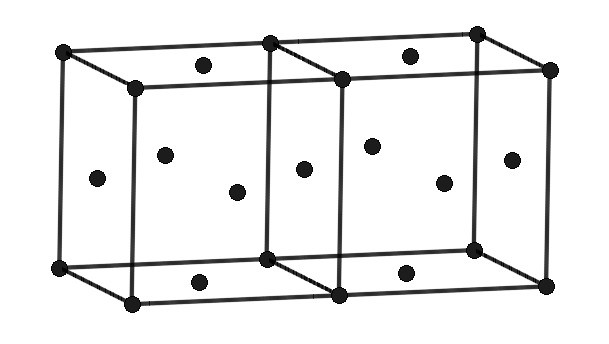}\hfill\includegraphics[width=.45\linewidth]{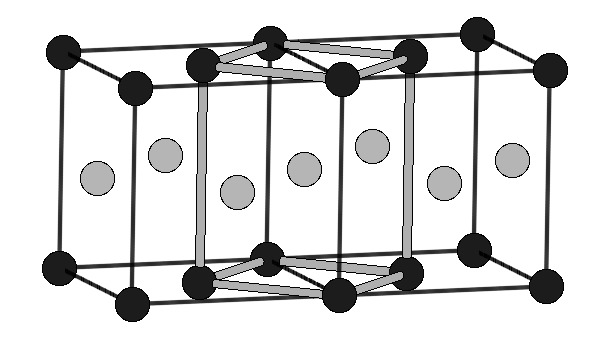}
\caption{ An example parent lattice (left) and a superstructure
  (right). The parent lattice is fcc and the superlattice is defined
  by the (doubled) unit cell outlined in gray. The two interior points
  of the superlattice are occupied by one black atom and one gray
  atom. The superlattice + atoms constitute a derivative
  structure. The superstructure of this example is that of
  CuAu.  \label{fig:superlattice}}
\end{figure}

The purpose of this paper is to present a general algorithm that generates a
formally complete list of two- or three-dimensional superstructures, and that
works for any parent lattice and for arbitrary $k$-nary systems (binary,
ternary, etc.). This algorithm is conceptually distinct from FWZ
and related implementations. Instead of using a geometrical, `smallest first'
approach to the enumeration,\cite{zarkevich:104203} it takes advantage of known
group theoretical properties of integer matrices. The algorithm is orders of
magnitude faster than FWZ, more general, and formally complete. A Fortran 95
implementation of the algorithm is included with this paper as supplementary
material.

\begin{figure*}
\noindent
\parbox{0.35\linewidth}{
  \caption{The first 17 binary structures derived from the fcc lattice. All have
    4 atoms/cell or less.  Structures shown with a green plane can be
    characterized as a stacking of pure A and B atomic layers. For example the
    L1$_0$ structure (upper left) is an alternating (A$_1$B$_1$) sequence of
    layers stacked in the [001] direction. All of the 2- and 3-atoms/cell
    structures have physical manifestations. Of the 4-atoms/cell structures only
    four have physical manifestations. Three of the others (yellow
    backgrounds) have been predicted to exist\cite{stefano_monster} but not yet
    observed. The other five (purple backgrounds) have never been observed or
    predicted to exist in any system.  \label{fig:17}}} \hfill
\parbox{0.60\linewidth}{\includegraphics[width=\linewidth]{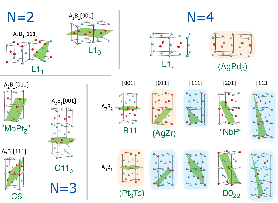}}
\end{figure*}

Mathematically, we can describe the purpose of the algorithm as this: for a
given parent lattice, enumerate all possible superlattices and all rotationally-
and translationally-unique `colorings' or labelings of each superlattice. In
presenting the algorithm in the next section, we shall refer to superlattices
and labelings rather than referring to crystal structures or atomic sites.

\section{Enumerating all derivative structures}
\label{sec:enum}
Here is a brief outline of the algorithm.
\begin{enumerate}
  \item For each superlattice of size $n$, generate all Hermite Normal Form (HNF)
    matrices.\cite{Santoro:a09671,Santoro:a08677} (In what follows, we refer to
    $n$ as the \emph{index} of the superlattice.)
  \item Use the symmetry of the parent lattice to remove rotationally-equivalent
    superlattices, thus shrinking the list of HNF matrices.\label{reduceHNFs}
  \item For each superlattice index $n$, find the Smith Normal Form (SNF) of each
    HNF in the list, and:\label{alg:SNF}
    \begin{enumerate}
    \item Generate a list of possible labelings (atomic configurations) for each
      SNF, essentially a list of all $k^n$ numbers in a base $k$, $n$-digit
      system.\label{knary} For the labels, we use the first $k$ letters of the
      alphabet, $a,b,\cdots$.
    \item Remove incomplete labelings where each of the $k$ labels
      ($a,b,\cdots$) does not appear at least once.
    \item Remove labelings that are equivalent under translation of the parent
      lattice vectors. This reduces the list of labelings by a factor of
      $\approx n$.
      \item Remove labelings that are equivalent under an exchange of labels,
        i.e., $a\rightleftharpoons b$, so that, e.g., the labeling $aabbaa$ is
        removed from the list because it is equivalent to $bbaabb$.
      \item Remove labelings that are super-periodic, i.e., labelings that
        correspond to a non-primitive superstructure. This can be done without
        using the geometry of the superlattice.
      \end{enumerate}
    \item For each HNF, remove labelings that are permuted by symmetry
      operations (of the parent lattice) that leave the superlattice
      fixed.\label{alg:labrot}
\end{enumerate}

\begin{figure}
\includegraphics[width=\linewidth]{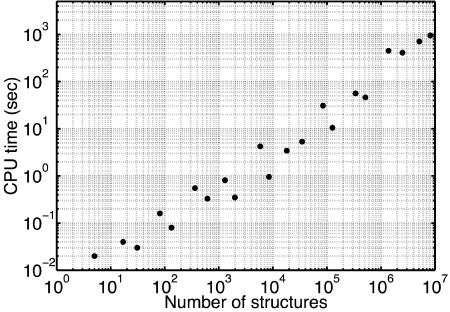}
  \caption{Required CPU time as a function of the number of unique derivative
    superstructures found. The scaling is
    linear, the best possible scaling for this type of problem.
  \label{fig:scaling}}
\end{figure}

An important feature of the algorithm is that the the list of possible
labelings, generated in step \ref{knary}, form a minimal hash table
with a perfect hash function. Eliminating \emph{all} duplicate
labelings in a list of $N$ can be accomplished in $O(N)$
time. Coupled with the group theoretical approach, this results in an
extremely efficient algorithm, orders of magnitude faster than FWZ,
and which scales \emph{linearly}. That is, the time to find $N$ unique
structures scales linearly with $N$, regardless of the size of
$N$. This linear scaling is shown for the case of binary
superstructures of an fcc parent lattice in Fig.~\ref{fig:scaling}.

\subsection{Generating all superlattices}
Given a `parent' cell (any lattice), the first step in finding all derivative
structures of that cell is to enumerate all derivative
super\emph{lattices}. Consider the transformation $\mathbb{B=AH}$ where
$\mathbb{A}$ is the basis for the original lattice (the basis vectors listed
column-wise), $\mathbb{H}$ is a matrix with all integer elements, and
$\mathbb{B}$ is the matrix of the transformed lattice. If the determinant of the
transformation matrix, $|\mathbb{H}|$, is $\pm1$, then $\mathbb{H}$ is merely a
change of basis that leaves the lattice represented by $\mathbb{A}$ unchanged.
Matrices $\mathbb{A}$ and $\mathbb{B}$ are merely two different choices of basis
for the same lattice.  On the other hand, if the elements of $\mathbb{H}$ are
all integers but the determinant of $\mathbb{H}$ is 2, say, then the lattice of
$\mathbb{B}$ is a superlattice\cite{Santoro:a08677}
of 
(i.e., a subgroup) of the lattice defined by $\mathbb{A}$, but with twice the
volume of the original (parent) lattice.

Two different matrices, $\mathbb{H}_1$ and $\mathbb{H}_2$, with the same
determinant, will generate different bases for the \emph{same lattice} if and
only if $\mathbb{H}_1$ can be reduced to $\mathbb{H}_2$ by elementary integer
column operations.  The canonical form for such operations is lower triangular
Hermite Normal Form (HNF).  Thus, if we use only matrices $\mathbb{H}$ which are
in HNF, we will produce exactly one representation of each
superlattice.\cite{Santoro:a09671,Santoro:a08677} In three dimensions, the
lower-triangular Hermite Normal Form is
\begin{equation}
\left(\begin{array}{ccc}
a & 0 & 0\\
b & c & 0\\
d & e & f\\
\end{array}\right)
\qquad\textrm{with}\qquad
0\leq b<c,\quad0\leq d,e < f.
\label{eq:conds}
\end{equation}
In this form, the product of the integers on the diagonal alone, $a\times
c\times f$, fixes the determinant. Again, we refer to the superlattice size, or
the determinant, as the index $n$. Generating all HNF matrices of a given
index can be done then by finding each unique triplet, $acf=|\mathbb{H}|$,
and then generating all values of $b$, $d$, and $e$ that obey the conditions in
(\ref{eq:conds}).

The algorithm for generating all possible HNF matrices of a given index
$|\mathbb{H}|$ is rather simple, comprising just two steps. In the first step,
find all possible diagonals: find all values $a$, $1\leq a \leq |\mathbb{H}|$,
that evenly divide $|\mathbb{H}|$; for each of these values, find all $c$,
$1\leq c \leq |\mathbb{H}|/a$, that evenly divide $|\mathbb{H}|/a$. For each
value of $c$, let $f=|\mathbb{H}|/(ac)$. For example, consider the case of
$|\mathbb{H}|=6$.  We execute two nested loops over the possible values of $a$
and $c$; each loop runs over all integers between 1 and the $|\mathbb{H}|$,
testing the above conditions at each iteration. The loops run from 1 to 6, and
the algorithm finds eight cases that meet the above conditions. They are:
\begin{equation*}
\begin{array}{c|cccccccc}
a &1&1&1&2&2&3&3&6\\
\hline
c &1&2&3&1&3&1&2&1\\
\hline
f &6&3&2&3&1&2&1&1\\
\end{array}
\end{equation*}

The set of $acf$ triplets generated during this first step comprises all
possible diagonals of the HNF matrices for the case of $n=|\mathbb{H}|=6$.  The
second step, generating each set of values of $b$, $d$, and $e$ for each
diagonal (set of $acf$ triplets), can be accomplished simply by three nested
loops that start at zero and terminate at $b<c$ and $d,e<f$.

As an example of both steps, consider the case where the index is
merely double that of the original lattice, i.e., where $|\mathbb{H}|=2$. The
factors of 2 are just the set $\{1,2\}$, so the first step finds only three cases:
(2,1,1), (1,2,1), and (1,1,2). Then, generating the off-diagonal terms for each
of these three cases, we find seven HNF matrices:
\begin{equation*}
\begin{array}{cccccc}
\textrm{case 1:}&
\left(\begin{array}{ccc}
2 & 0 & 0\\
0 & 1 & 0\\
0 & 0 & 1\\
\end{array}\right)&\quad&\textrm{case 2:}&
\left(\begin{array}{ccc}
1 & 0 & 0\\
0 & 2 & 0\\
0 & 0 & 1\\
\end{array}\right)&
\left(\begin{array}{ccc}
1 & 0 & 0\\
1 & 2 & 0\\
0 & 0 & 1\\
\end{array}\right)\\
\end{array}\\
\end{equation*}
\begin{equation*}
\begin{array}{ccccc}
\textrm{case 3:}&
\left(\begin{array}{ccc}
1 & 0 & 0\\
0 & 1 & 0\\
0 & 0 & 2\\
\end{array}\right)&
\left(\begin{array}{ccc}
1 & 0 & 0\\
0 & 1 & 0\\
0 & 1 & 2\\
\end{array}\right)&
\left(\begin{array}{ccc}
1 & 0 & 0\\
0 & 1 & 0\\
1 & 0 & 2\\
\end{array}\right)&
\left(\begin{array}{ccc}
1 & 0 & 0\\
0 & 1 & 0\\
1 & 1 & 2\\
\end{array}\right)\\
\end{array}
\end{equation*}

For increasing index, $n=|\mathbb{H}|=1,2,3,\cdots$, the number of HNF
matrices generates an interesting sequence: 1, 7, 13, 35, 31, 91, $\cdots$. This
sequence appears in Sloane's database\cite{sloane} as A001001. The
closed-form expression for $n$-th term in the series is
\begin{equation}
\sum_{d\big\vert n}d\>\sigma(d)=
\prod_{i=1}^k\left(\frac{(p_i^{e_i+2}-1)(p_i^{e_i+1}-1)}{(p_i-1)^2(p_i+1)}\right),
\label{eq:hnfcount}
\end{equation}
where $\sigma$ is the sum of divisors function, and
where the $p_i$ and $e_i$ are the prime factors and powers of $n$:
$n=p_1^{e_1}\cdot p_2^{e_2}\cdots p_k^{e_k}$. The sequence appears in the
crystallography literature\cite{Rutherford:he0009,Billiet_LeCoz:80} as well as several other
contexts.\cite{liskovets:2000,baake:97,stanley:99,sloanes:A001001}

Significantly, because we have an expression for the number of superlattices, the
implementation of the HNF-generating algorithm can be rigorously checked. Also
note that this step of the algorithm is independent of the choice of parent
lattice. 

\subsection{Reducing HNF list by parent lattice symmetry}
The set of HNF matrices defines the set of all derivative superlattices of a
parent cell via the transformation mentioned above, $\mathbb{B=AH}$. However,
not all of the superlattices in this set will be geometrically different. 
Some distinct lattices will be equivalent under symmetries of the parent
lattice, illustrated in the example below.

Such duplicate superstructures must be eliminated by
the algorithm. At the end of the algorithm we want all derivative structures to be
unique from a materials point of view. So we wish to exclude from the list any
superstructures that are related to others already in the list simply by a rotation, reflection
or change of basis. 

As an illustration, consider a two-dimensional parent lattice that is
square, that is, $\mathbb{A=I}$ (the two-by-two
identity matrix). There are three HNF matrices for which $|\mathbb{H}|=2$
and three corresponding superlattices, $\mathbb{B=AH=IH=H}$:

\parbox{.9\linewidth}{
\parbox{2cm}{$$
\left(\begin{array}{cc}
2 & 0\\
0 & 1\\
\end{array}\right)$$}
\hfill
\parbox{2cm}{
$$\left(\begin{array}{cc}
1 & 0\\
0 & 2\\
\end{array}\right)$$}
\hfill
\parbox{2cm}{
$$\left(\begin{array}{cc}
1 & 0 \\
1 & 2 \\
\end{array}\right)$$
}
\includegraphics[width=\linewidth]{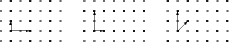}}\\ The parent lattice itself
is indicated by dots (filled and unfilled) while the superlattice is indicated
by filled dots. The vectors defined by the matrices are shown as arrows.
The first two lattices are clearly equivalent under a 90$^\circ$ rotation, one
of the eight symmetry operations of a square lattice.

To enumerate the distinct superlattices of a given index $n$ then, we
must check that each new superlattice that is added to the list is not a rotated
duplicate of a previous superlattice.  More precisely, we must check
that each new basis $\mathbb{B}_i$  is not equivalent, under change-of-basis,
to some symmetric image $\mathbb{RB}_j$ of a basis $\mathbb{B}_j$
already in
the list. In other words, we want to avoid the relation
$
\bb{B}_i=\bb{RB}_j\bb{H}
$
where $\bb{B}_i$ is a candidate superlattice, $\bb{R}$ is any of the rotations
of the parent lattice, $\bb{B}_j$ is a superlattice already in the list of
distinct superlattices, and $\bb{H}$ is any unimodular matrix of integers. 
(Since $\bb{B}_i$ and $\bb{B}_j$ have the same determinant, we will
only need to check that $\bb{B}_j^{-1}\bb{R}^{-1}\bb{B}_i$ is a 
matrix of integers.)

For the case of cubic symmetry, the seven superlattices for the $\mathbb{H}=2$
case mentioned above reduce to only two symmetrically-distinct
superlattices. The corresponding derivative superstructures are L1$_0$ and
L1$_1$, both well-known structures in intermetallic compounds. The fact that
these are the only two 2-atom/cell fcc structures is not coincidence or an
accident of chemistry; no other 2-atom/cell structures are \emph{possible
  geometrically}. The hierarchy of physically-observed structures uncovered for
fcc and bcc lattices as the index is increased is discussed in
Refs.~\onlinecite{missing,L1_3}. Additional applications can be found in
Sec.~\ref{sec:apps}.

\subsection{Find the unique labelings for all superlattices}
\subsubsection{Generate all possible labelings}\label{sec:all_labelings}
For each HNF, each superlattice, we start by generating all possible
labelings of that superlattice. In other words, given $k$ colors
(types of atoms), represented by the labels $a,b,\cdots$, we generate
all possible ways of labeling (coloring) the superlattice.  Each HNF
matrix of determinant size $n$ represents a superlattice with $n$
interior points to be decorated. If the number of colors is $k$, then
the list of all possible labelings is easily represented by the list
of all $n$-digit, base-$k$ numbers.  So, from a combinatorial point of
view, there are $k^n$ distinct labelings. For example, in the case of a
binary system ($k=2$) with 4 interior points (index $n=4$), there are
$2^4=16$ possible labelings (see Table~\ref{tab:two4ex}).
\definecolor{incomplete}{cmyk}{0,0,1,0}
\definecolor{transdup}{cmyk}{0,.5,0,0}
\definecolor{labperm}{cmyk}{.5,0,0,0}
\definecolor{nonprim}{cmyk}{.5,0,.5,0}

\begin{table}
\begin{equation*}
\begin{array}{cccc}
\colorbox{incomplete}{aaaa}&\colorbox{transdup}{abaa}&\colorbox{transdup}{baaa}&\colorbox{transdup}{bbaa}\\
aaab&\colorbox{nonprim}{abab}&\colorbox{transdup}{baab}&\colorbox{labperm}{bbab}\\
\colorbox{transdup}{aaba}&\colorbox{transdup}{abba}&\colorbox{transdup}{baba}&\colorbox{labperm}{bbba}\\
aabb&\colorbox{labperm}{abbb}&\colorbox{labperm}{babb}&\colorbox{incomplete}{bbbb}\\
\end{array}
\end{equation*}
\caption{An example labeling for the binary case, $k=2$, with 4
  interior points, $n=4$, in the superlattice. There are $k^n=2^4=16$
  distinct labelings, but the colored labelings represent incomplete
  or duplicate superstructures. Yellow labelings are incomplete (not
  all labels are present), purple are translation duplicates, blue are
  label-exchange duplicates, and green are super-periodic
  labelings. \label{tab:two4ex}}
\end{table}

\subsubsection{Concept of eliminating duplicate labelings}
The rest of the algorithm deals with just one conceptual issue---given
the $k^n$ labelings/colorings of the superlattice, eliminate the
duplicates. In the FWZ algorithm and its extended implementations,
duplicate structures are eliminated by directly comparing\cite{note4}
one candidate structure to another geometrically, necessitating an
expensive $O(N^2)$ search. We eliminate the duplicates via group
theory rather than checking the structures themselves. Although this
approach is more abstract than the geometric approach, it is much more
efficient---eliminating the duplicates in a list becomes a strictly
$O(N)$ procedure.

\begin{figure}
\includegraphics[width=\linewidth]{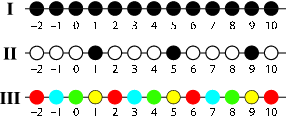}
\caption{ (I) One-dimensional example of a parent lattice, (II) a derivative
  superlattice (index $n=4$), (III) one possible superlattice labeling.\label{1D}}
\end{figure}

\emph{One-dimensional example:}
We start
with a simple illustration and then discuss the essential group theoretical
concepts in the context of that example. Consider the one-dimensional case of
Fig.~\ref{1D}. The first line (I) is a parent lattice, an infinite collection of
equally-spaced points, identified with the set of integers, denoted ${\bf Z}$. 
The second line (II) is a superlattice, a subset of the
parent lattice (every fourth point; those colored black). The third line (III) is a
super\emph{structure}, a `labeling' or `coloring' of the parent lattice that
has the same periodicity as the superlattice. The points of the lattice play the
role of positions in a crystal, and the colors/labels play the role of
atoms placed at those positions.

\begin{figure}
\includegraphics[width=\linewidth]{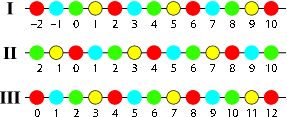}
\caption{ Three labelings of the superlattice of Fig.~\ref{1D}. Lines II and III
  are identical to line I except the colors (labels) are shifted. In line II the
  colors have been shifted two units to the right. In line III, the shift has
  been effected by translating the lattice two units to the left.\label{1D_shift}}
\end{figure}

There are labelings that are distinct yet
physically equivalent, as shown in Fig.~\ref{1D_shift}.  Note that line (II) is obtained
from line (I) by shifting the colors two units to the right, and line (III) is obtained
from line (I) by shifting the numbers two units to the left---with the same result.
Lines (II) and (III) are the same labeling, obtained in different ways from
(I). Both are physically equivalent to (I).  The fact that we can obtain such a shifted
labeling either by shifting the numbers or by shifting the labels explains why
we can do much of our equivalence checking within a finite group, instead of
geometrically within an infinite lattice.  By this method, we will identify these
equivalent labelings and remove them from the original list of $k^n$ labelings.

We may illustrate the group theory approach using Fig.~\ref{1D}. The parent
lattice (I) is the set of integers ${\bf Z}$, which is a \emph{group} under the
addition operation. We refer to this group as $L$. The superlattice is the set
of multiples of 4, denoted $4{\bf Z}$. We refer to this \emph{subgroup} of $L$
as $S$.  We label the parent lattice $L$ in a manner which is periodic with
respect to the superlattice $S$, and note that if two points differ by an
element of the superlattice, they must receive the same label.  We use colors as
labels in line (III) of Fig.~\ref{1D} and note that every fourth point has the
same color.

Notice that the green points are our superlattice are $4{\bf Z}$, and the yellow
points are a copy of $4{\bf Z}$, but \emph{translated one unit to the right}.
Thus, we may denote the latter set (the yellow points) by the set $1+4{\bf Z}$.
Similarly, the red points are the set $2+4{\bf Z}$, and the blue points are
$3+4{\bf Z}$.  These four sets, $4{\bf Z},\, 1+4{\bf Z},\, 2+4{\bf Z},\, 3+4{\bf
  Z}$, are mutually disjoint (they don't overlap), and their union is the entire
parent lattice $L$.  They are \emph{translations} of $S$; and thus are the
\emph{cosets} of the subgroup $S$. This means we can use them to form a new
group, called the \emph{quotient group} $G=L/S$ (see Table~\ref{tab:cayley}).
This new group is \emph{finite}, having only four elements. For notational
convenience we shall also refer to these 4 elements of $G$ as (0,1,2,3). We need only
label the four elements of our quotient group in order to label the \emph{entire
  parent lattice}.

Suppose we wish to translate a labeling (in order to identify and eliminate
equivalent structures).  As shown in line (I) of Fig.~\ref{1D_shift}
we have labeled the elements of the quotient group as follows (using $g$, $y$, $r$
and $b$ for the colors): $$
\begin{array}{lccc}
0:& \quad 4{\bf Z}&\rightarrow&  g\\
1:& \quad1+4{\bf Z}&\rightarrow& y\\
2:& \quad2+4{\bf Z}&\rightarrow& r\\
3:& \quad3+4{\bf Z}&\rightarrow& b.
\end{array}$$
In Fig.~\ref{1D_shift}, we see that translating the labels by two is the 
same as simply adding 2 to each coordinate, thus: 
$$\begin{array}{rcl}
4{\bf Z} &\rightarrow&  2+4{\bf Z}\\ 1+4{\bf Z} &\rightarrow& 3+4{\bf Z}\\
2+4{\bf Z}& \rightarrow& 4+4{\bf Z}=4{\bf Z}\\ 3+4{\bf Z} &\rightarrow& 
5+4{\bf Z}=1+4{\bf Z} \end{array}\qquad
\begin{array}{rcl}
0&\rightarrow&2\\1&\rightarrow&3\\2&\rightarrow&0\\3&\rightarrow&1\\
\end{array}
$$
The effect is the same as if we had assigned the colors differently:
$$\begin{array}{lccc}
0:& \quad 4{\bf Z} &\rightarrow& r\\
1:& \quad1+4{\bf Z}&\rightarrow& b\\
2:& \quad2+4{\bf Z}&\rightarrow& g\\
3:& \quad3+4{\bf Z}&\rightarrow& y.
\end{array}$$
Translating the lattice by adding $+2$ to every
point (moving the origin by 2 units) has the same effect on the labeling as if
we had merely labeled the four elements of the quotient group, and
then added $+2$ to every element of the group, producing the permutation
$0\rightarrow 2$, $1\rightarrow 3$, $2\rightarrow 0$ and $3\rightarrow 1$,
denoted (2,0,3,1).

\begin{table}
\parbox{.7\linewidth}{
\begin{tabular}{c|cccc}
             & $\mathbf{Z}$ &$1+\mathbf{Z}$ &$2+\mathbf{Z}$ &$3+\mathbf{Z}$\\
\hline
$\mathbf{Z}$  &$\mathbf{Z}$  &$1+\mathbf{Z}$ &$2+\mathbf{Z}$ &$3+\mathbf{Z}$  \\
$1+\mathbf{Z}$&$1+\mathbf{Z}$&$2+\mathbf{Z}$&$3+\mathbf{Z}$&$\mathbf{Z}$\\
$2+\mathbf{Z}$&$2+\mathbf{Z}$&$3+\mathbf{Z}$&$\mathbf{Z}$&$1+\mathbf{Z}$\\
$3+\mathbf{Z}$&$3+\mathbf{Z}$&$\mathbf{Z}$&$1+\mathbf{Z}$&$2+\mathbf{Z}$\\
\end{tabular}
}\hfill\parbox{.3\linewidth}{
\begin{tabular}{c|cccc}
             & $0$ &$1$ &$2$ &$3$\\
\hline
$0$  &$0$  &$1$ &$2$ &$3$  \\
$1$&$1$&$2$&$3$&$0$\\
$2$&$2$&$3$&$0$&$1$\\
$3$&$3$&$0$&$1$&$2$\\
\end{tabular}
}
\caption{ Two representations of the Cayley table for the quotient group $G$
  (cosets of the subgroup 
  $S$). On the right, the elements of the group have been denoted (0,1,2,3) for
  notational convenience. \label{tab:cayley}}
\end{table}

Instead of determining that two labelings of the (infinite) lattice are
equivalent by translation, we may simply check that the corresponding labelings
of our finite quotient group $G={\bf Z}_4$ are equivalent. We do this by just
adding a fixed element to every element in the group, effecting a permutation of
the cyclic group ${\bf Z}_4$. This idea---of labeling the quotient group instead
of the lattice elements, and checking equivalence within the group instead of by
translating the lattice---may seem unduly abstract and unnecessary in one
dimension, but it becomes much more efficient and crucial in higher dimensions,
as we now show.

\emph{Application to higher dimensions:} In any dimension, we have a parent
lattice $L$, and a superlattice $S$ which is a {\it subgroup} of $L$.  Labeling
$L$ in a manner which is periodic with respect to $S$ is equivalent to merely
labeling the elements of the quotient group $G=L/S$.  Note that even though $L$
and $S$ are infinite sets, their quotient group is always a {\it finite} group
with the same number of elements as the superlattice index $n$.  Again, we check
for equivalence by doing operations within the group instead of by lattice
translation.

\begin{figure}
\includegraphics[width=\linewidth]{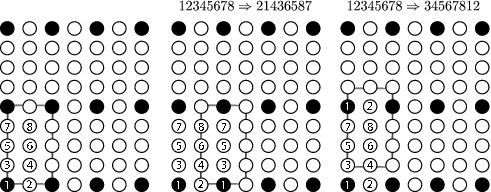}
\caption{ A superlattice whose SNF is non-trivial (non-cyclic). Because the quotient group of
  the superlattice is non-cyclic, translations of the lattice are not cyclic
  permutations. Instead the sites are permuted in groups of two or four with each
  translation (rather than permuting as a single group of 8). In the figure, three different pictures of
  the same superlattice are shown. In each case a different origin is
  chosen. In the second case, the translation permutes two groups of 4 sites, in
  the third, 4 pairs are permuted. \label{fig:2D_cycle}}
\end{figure}

The key to this approach is the Smith normal form (SNF). The SNF is useful
because it provides the quotient group directly as follows. Recall that if
$\mathbb{A}$ is a basis for $L$, then the distinct lattices of index $n$ are
uniquely characterized by bases $\mathbb{B=AH}$, where $\mathbb{H}$ is a matrix
of determinant $n$ in HNF.  If $S$ is given by one such basis
$\mathbb{B}_1=\mathbb{AH}_1$, then the quotient group $G=L/S$ can be found by
converting the matrix $\mathbb{H}_1$ into SNF (which is a diagonal matrix with
certain special properties; see Appendix). In higher dimensions the quotient
group may not be purely cyclic, but it is a direct sum of cyclic groups which
are given by the diagonal entries in the SNF matrix (see Fig.~\ref{fig:2D_cycle}). For example if the SNF
matrix is
$\begin{pmatrix} 2&0\\0&4\\
\end{pmatrix}$, then the quotient group $G=L/S$ is the direct sum ${\bf Z}_2\oplus {\bf Z}_4$. 

In relation to the algorithm, there are two important facts to note about the
SNF. (i) The SNF provides the quotient group directly, which in turn is the key
to implementing an $O(N)$ algorithm. (ii) The number of SNFs (and so quotient
groups) is small compared to the number of distinct lattices of index $n$ (see
Table~\ref{tab:snf}). This means that translation duplicates can be removed from
the $k^n$ list for hundreds or thousands of \emph{different} superlattices
\emph{simultaneously}. (The surprising geometric implications of this are
discussed in the appendix.) This reduces the running time by many orders of
magnitude.   

\begin{table}
\begin{tabular}{c|ccccccccccccccccccc}
$n$& 2& 3& 4& 5& 6& 7& 8& 9& 10& 11& 12& 13& 14& 15&16\\
\hline
HNFs &7& 13& 35& 31& 91& 57& 155& 130& 217& 133& 455& 183& 399& 403&651\\
\hline
SNFs &1 & 1 & 2 & 1 & 1 & 1 & 3 & 2 & 1 & 1 & 2 & 1 & 1 & 1  &4 \\
\end{tabular}  
\caption{Table showing the number of Hermite normal form (HNF) matrices and Smith
  normal form (SNF) matrices as a function of index $n$ (determinant size). The number
  of HNFs is a rapidly-increasing function of $n$ (see Eq.~\ref{eq:hnfcount}) whereas the
  number of SNFs grows very slowly.\label{tab:snf}} 
\end{table}
\
\subsubsection{Eliminating translation duplicates}
Because of its periodicity, the choice of origin of a superlattice is
arbitrary. A change in origin implies a permutation of the labels that
nonetheless defines the same superstructure (compare lines I and III in
Fig.~\ref{1D_shift}). As stated previously, by examining the quotient group instead of
directly comparing the structures, the duplicate labelings can be readily
identified. For example, consider the case for $n=4$. Adding each member to the
quotient group $\mathbf{Z}_4=(0,1,2,3)$ effects 4 permutations as follows:
\begin{center}
\begin{tabular}{clc}
member&\multicolumn{1}{c}{mapping}&permutation\\
\hline
0:&0$\rightarrow0$, 1$\rightarrow1$, 2$\rightarrow2$, 3$\rightarrow3$&(0,1,2,3)\\
1:&0$\rightarrow1\rightarrow2\rightarrow3\rightarrow0$&(1,2,3,0)\\
2:&0$\rightarrow2\rightarrow0$, 1$\rightarrow3\rightarrow1$&(2,0,3,1)\\
3:&0$\rightarrow3\rightarrow2\rightarrow1\rightarrow0$&(3,0,1,2)\\
\end{tabular}
\end{center}
If we take the 14 complete labelings of Table~\ref{tab:two4ex} and the three
non-trivial permutations above, we find that 10 are duplicates (colored
purple in Table~\ref{tab:two4ex}):\\
\begin{center}
\begin{tabular}{c|ccc}
(0,1,2,3)&(1,2,3,0)&(2,0,3,1)&(3,0,1,2)\\
&\multicolumn{3}{c}{duplicates}\\
\hline
aaab&aaba&abaa&baaa\\
aabb&abba&bbaa&baab\\
abab&baba&abab&baba\\
abbb&bbba&bbab&babb\\
\end{tabular}
\end{center}
Of the original $2^4=16$ labelings, two were discarded immediately
because they were incomplete. Of the remaining 14, 10 are translation
duplicates, leaving 4 that are translationally inequivalent (left
column above). 

\subsubsection{Remove `label-exchange' duplicates}
The next step in the algorithm is to remove labelings that are equivalent under
exchange of labels. Structurally, there is no difference between a superlattice
whose interior points are labeled $aaab$ versus $bbba$. Although the energy of
an isostructural compound with composition X$_3$Y$_1$ is different from one with
composition X$_1$Y$_3$, we only wish to include one entry in our list of
derivative superstructures because the full composition list can always be
recovered by making all possible label exchanges (i.e., $a\rightleftharpoons
b$). In the example above, four labelings were unique under translations:
\begin{center}
\begin{tabular}{c}
aaab\\
aabb\\
abab\\
abbb\\
\end{tabular}
\end{center}
But the first and the fourth are equivalent by exchanging
$a\leftrightharpoons b$ and applying the permutation (1,2,3,0). 

\begin{figure}
\includegraphics[width=\linewidth]{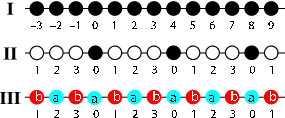}
\caption{ (color on-line) (I) One-dimensional example of a parent lattice, (II)
  a derivative superlattice (index $n=4$), (III) a super-periodic (or
  non-primitive) labeling. Although the index of the superlattice is $n=4$, the
  structure can be represented by a superlattice labeling of period 2 instead
  of 4. The superstructure of line III would have been found as an index $n=2$
  derivative structure and is therefore a duplicate.\label{fig:nonprimitive}}
\end{figure}

\subsubsection{Remove super-periodic labelings (non-primitive structures)}
At this point of the algorithm, many of the duplicate labelings have
been removed from the original $k^n$ list. But there are still more
duplicates to remove.  Some of the labelings in the list will
represent superstructures that are not primitive. In other words, the
labelings will be super-periodic---they will have periods shorter
scale than the superlattice.\cite{note_rutherford} 


The super-periodic duplicates are easily identified because they are identical
under at least one permutation. The quotient group $G$ dictates a set of
permutations under which the labelings are duplicate. One of these permutations
will leave the labeling unchanged if it is super-periodic. For example,
continuing the example above, three unique labelings are still in the list:
$aaab$, $aabb$, and $abab$. One of the permutations of the quotient group
$G=\mathbf{Z}_4$ is (2,0,3,1). Under this permutation, the labeling $abab$ is
unchanged. Thus it is super-periodic, as depicted in
Fig.~\ref{fig:nonprimitive}. It is a duplicate in the sense that the algorithm
would have already enumerated this structure with the index $n=2$ structures.

\subsubsection{For each HNF: remove `label-rotation' duplicates}
The previous three steps of the algorithm yield a list of distinct labelings
for each SNF of index $n$. Three kinds of the duplicate labelings, translation
duplicates and label-exchange duplicates and super-periodic duplicates, have
already been removed. One kind of duplicate remains, however, and these are
eliminated in the current step.

This step removes labelings which are permuted by the rotations of the parent
lattice. Whereas the preceding steps were applied to generate a list of unique
labelings \emph{for each SNF}, the current step must be applied to \emph{each
  HNF}. In other words, this step must be applied to the surviving labelings
\emph{separately} for each superlattice. 

\begin{figure}
\includegraphics[width=.5\linewidth]{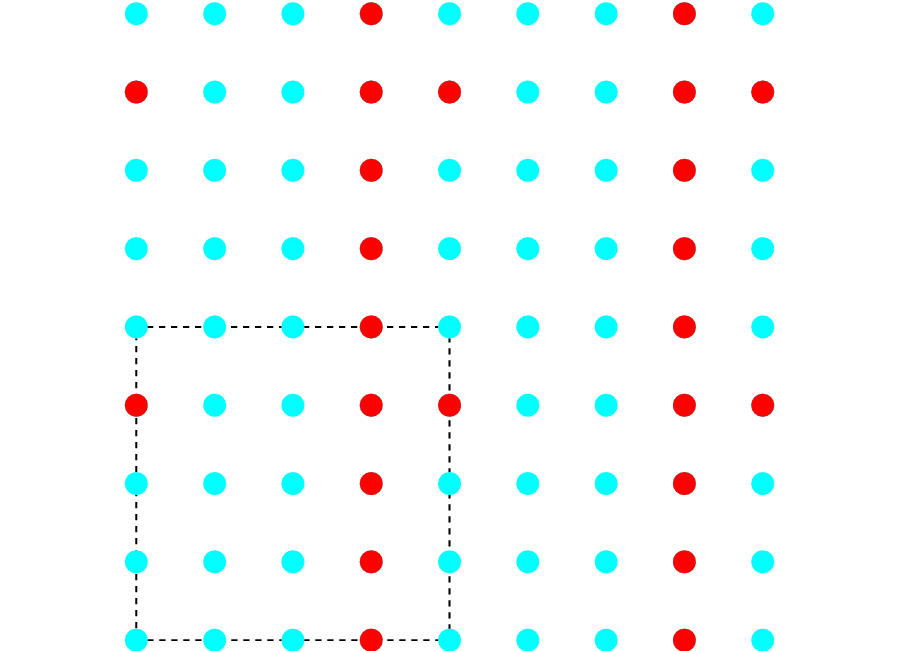}\hfill\includegraphics[width=.5\linewidth]{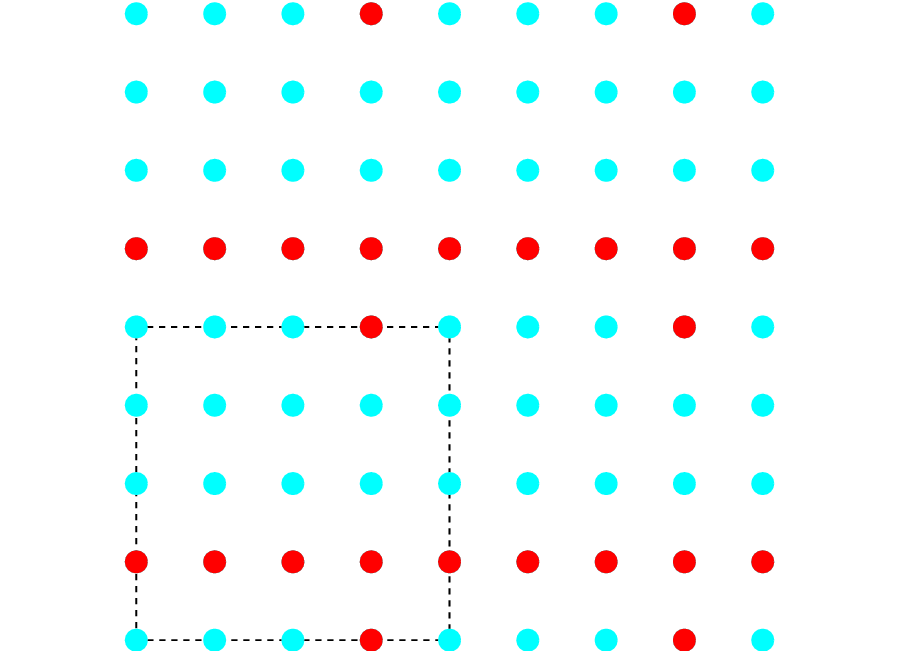}
\caption{ (color on-line) Two identical $4\times4$ superlattices (dotted lines) with
  rotationally-equivalent labelings (red and blue circles). Although the superlattices
  themselves are unchanged, a $90^\circ$ rotation applied to the left
  labeling yields that shown on the right. Thus the second is a
  duplicate of the first and should be removed from the list of labelings. \label{fig:labrot}}
\end{figure}

Superlattices which are not fixed by
rotations of the parent lattice were already eliminated as duplicates in step
\ref{reduceHNFs} of the algorithm. But rotations which leave the superlattice
unchanged may still permute the labeling itself. Such permutations are
physically equivalent (merely rotated with respect to one another). So any two
labelings which are equivalent under rotations that fix the superlattice are
duplicate and one must be removed from the list. Figure~\ref{fig:labrot}
illustrates the situation in two dimensions.

Here again, the group theory approach and the SNF makes the search extremely
efficient.  Label rotation duplicates can be identified easily using the
properties of the quotient group and the SNF transformation. The row and column
operations required to transform the HNF matrix of a superlattice into its SNF
can be represented by two integer transform matrices, $\mathbb{L}$ and
$\mathbb{R}$, so that $\mathbb{LHR=S}$, where $\mathbb{S}$ is the SNF.  This
step of the algorithm is implemented using the left transformation matrix
$\mathbb{L}$.

Let $\mathbb{G}$ be a $3\times n$ matrix where each member of the quotient group is
represented as a column\cite{imagegroup} in $\mathbb{G}$, and let $\mathbb{R}$
be one of the rotations that fixes the superlattice. Then the permutation of the
labels (which is the same as the permutations of the quotient group) enacted by the
rotation $\mathbb{R}$ is given by:
\begin{equation}
\mathbb{G}'=\mathbb{LA}^{-1}\mathbb{R}(\mathbb{LA}^{-1})^{-1}\mathbb{G}
\end{equation}
where columns of $\mathbb{A}$ are the lattice vectors of the
\emph{parent lattice}, and  $\mathbb{L}$ is the left transformation matrix for
the SNF. 

The power of this expression is that it allows the label-rotation duplicates to
be identified by working entirely within the quotient group, without requiring
any explicit reference to the geometry of the superlattice. Thus, as in the
other steps, duplicates labelings can be eliminated in a time proportional only
to the number of labelings in the list.

\section{Examples}
In this section, we give several example derivative structure lists
enumerated by the algorithm. We discuss the symmetry reduction of the structure
lists and then give results for several cases. First, we compare the fcc/bcc
binary list to that generated by the FWZ algorithm. We also show the
small-unit-cell binary structures for a simple-cubic parent lattice and the
small-unit-cell \emph{ternary} structures for an fcc parent lattice.

\begin{figure}
\includegraphics[width=.7\linewidth]{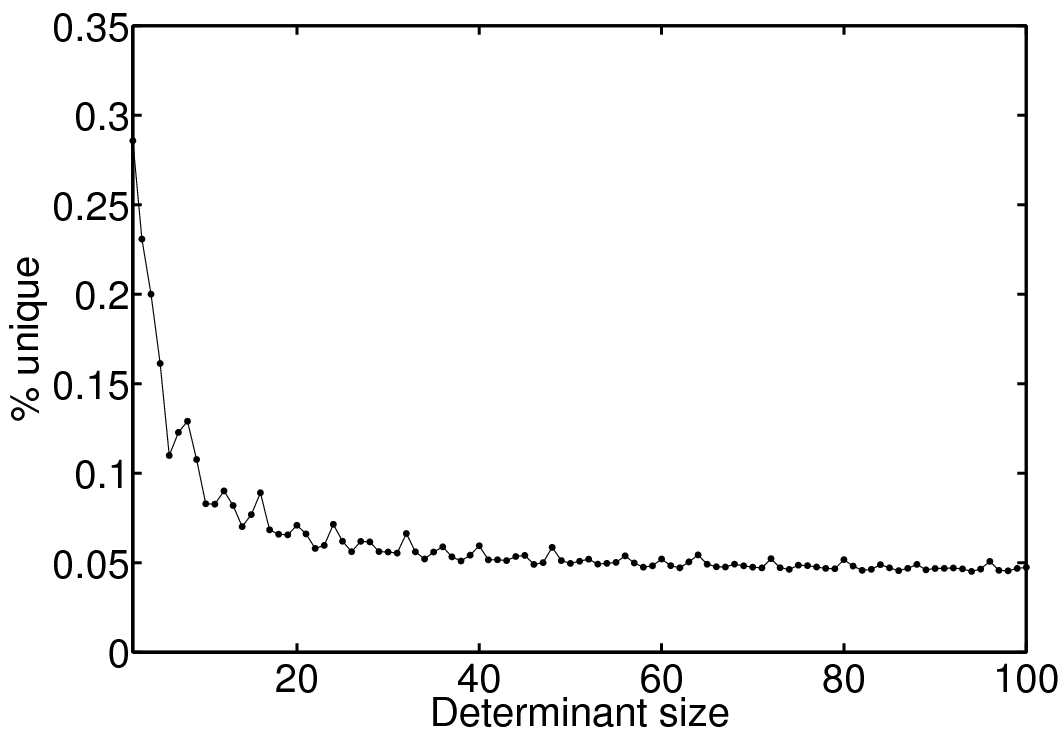}
\caption{Symmetrically unique HNF matrices as a function of the index $n$
  (determinant size) for an fcc parent lattice. Asymptotically, the fraction of
  unique superlattices approaches $2/N=1/24\approx5$\%, where $N=48$ is the
  number of symmetry operations of the fcc parent lattice. \label{fig:dups}}
\end{figure}

\subsection{Symmetry reduction of superlattices}
\begin{figure}
\includegraphics[width=\linewidth]{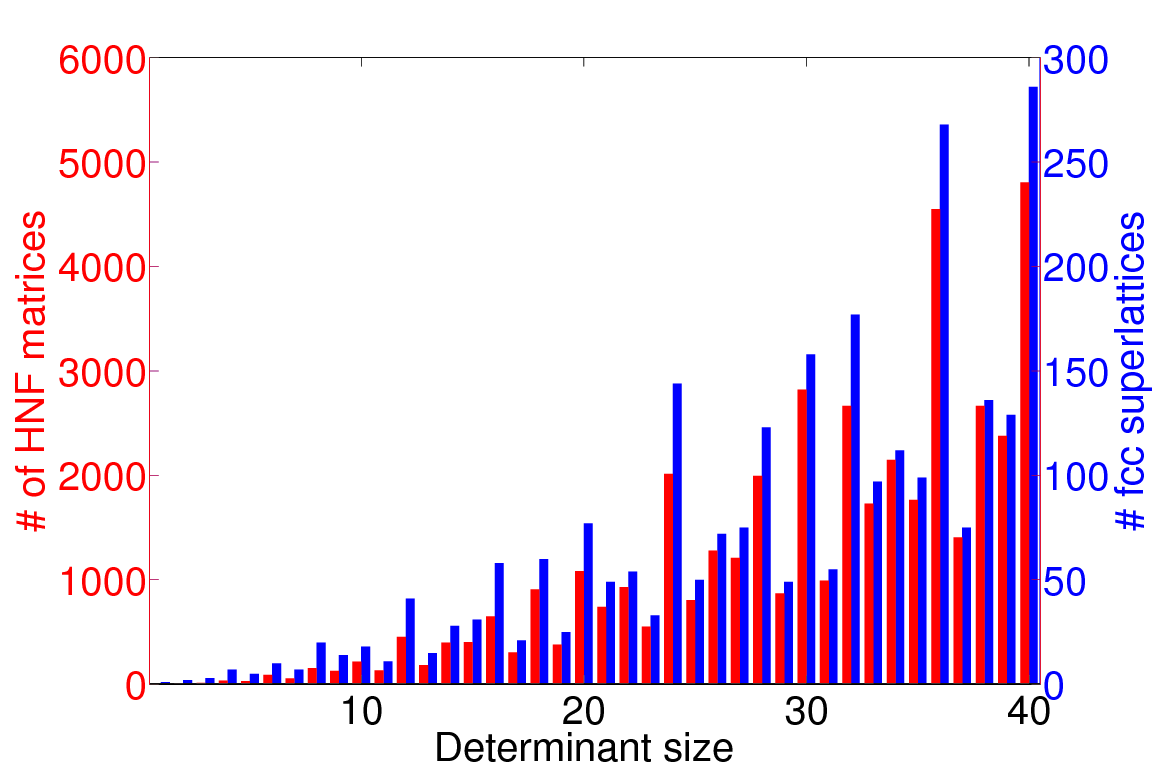} 
\caption{(color on-line) Left axis (red): Number of HNF matrices as a
  function of determinant size. Right axis (blue): Number of inequivalent fcc
  superlattices as a function of volume. \label{fig:hnfs}}
\end{figure}

In step B of the algorithm, the complete list of HNF matrices is reduced to
those that are unique under the symmetry operations of the parent
lattice. Asymptotically, the factor by which the list is reduced is one half the
order of the rotation group of the parent lattice. For example, for cubic parent
lattices (simple cubic, face-centered-, or body-centered-cubic), the point group
contains 48 rotations (proper and improper). For superlattices with large index
$n$, the number of HNFs is reduced by a factor of $48/2=24$. Because every
lattice is symmetric under inversion, only the proper rotations (i.e., not
reflections) need be considered in the reduction (thus the factor of
1/2). Figure~\ref{fig:dups} shows the fraction of symmetrically distinct
superlattices for determinant sizes up to 100, while Fig.~\ref{fig:hnfs} shows
the actual number of fcc-based superlattices compared to the total number of
distinct HNF matrices.

For an fcc or bcc parent lattice (the numbers are the same), the number of
unique lattices as a function of index $n$ (cell size) is equivalent to
the Sloane sequence A045790.\cite{sloanes:A045790} For the sequences 
generated for other parent lattices, which accordingly have a different symmetry
group, there are no known number-theoretic connections. Surprisingly,
this is even true for the simple cubic lattice. For the simple cubic lattice,
the sequence is identical to the fcc/bcc one for \emph{odd} values of the index
$n$ but larger for the even values. (See Table~\ref{tab:unique_sls}.)

\begin{table}
\begin{tabular*}{0.75\linewidth}{@{\extracolsep{\fill}}c|cccc}
index&\multicolumn{4}{c}{\# of superlattices}\\
$n$\qquad & fcc/bcc & sc & hex & tetragonal \\
\hline
2 & 2  &      3  &       3    &   5 \\
3 & 3  &      3  &       5    &   5 \\
4 & 7  &      9  &      11    &  17 \\
5 & 5  &      5  &       7    &   9 \\
6 & 10 &     13  &      19    &  29 \\
7 & 7  &      7  &      11    &  13 \\
8 & 20 &     24  &      34    &  51 \\
9 & 14 &     14  &      23    &  28 \\
10& 18 &     23  &      33    &  53 \\
\end{tabular*}  
\caption{ Number of symmetrically unique superlattices (HNFs) for several
  different parent lattices as a function of the index size $n$. Note that
  fcc/bcc are the same as simple cubic (sc) only for the odd values of
  $n$, and always smaller for the even values. Hexagonal (hex) and simple
  tetragonal parent lattices have more unique lattices than the cubic systems
  because of their lower symmetry.\label{tab:unique_sls}}  
\end{table}

\subsection{Number of structures of different parent lattices}
The number of super\emph{structures} increases much faster than the number of
super\emph{lattices} as a function of $n$. In general, each superlattice has
many different unique labelings. Table~\ref{tab:tot_enum} shows the number of
fcc/bcc derivative structures as a function of $n$. The FWZ begins to undercount
(as expected) at $n=15$ but the FWZ count is probably sufficient for
applications where it was used. Our algorithm is formally complete and does not
undercount. 

\begin{table}
\begin{tabular*}{0.95\linewidth}{@{\extracolsep{\fill}}rrr|rrr}
$n$&structures&cumulative&$n$&structures&cumulative\\
\hline
   2   &            2  &          2  &          13 &           2\,624 &        8\,480 \\
   3   &            3  &          5  &          14 &           9\,628   &     18\,108\\
   4   &           12  &         17  &          15 &          16\,584   &     34\,692\\
   5   &           14  &         31  &          16 &          49\,764   &     84\,456\\
   6   &           50  &         81  &          17 &          42\,135   &    126\,591\\
   7   &           52  &        133  &          18 &         212\,612   &    339\,203\\
   8   &          229  &        362  &          19 &         174\,104   &    513\,307\\
   9   &          252  &        614  &          20 &         867\,893   &   1\,381\,200\\
  10   &          685  &       1\,299  &        21 &        1\,120\,708   & 2\,501\,908\\
  11   &          682  &       1\,981  &        22 &        2\,628\,180   & 5\,130\,088\\
  12   &         3\,875  &     5\,856  &        23 &        3\,042\,732   & 8\,172\,820\\
\end{tabular*}  
 \caption{ Number of unique fcc derivative structures as a function of the index $n$.
 The second and fifth columns show the number of unique structures for \emph{each} $n$,
 while the third and sixth columns show the cumulative total.\label{tab:tot_enum}}  
\end{table}

\begin{table}
\begin{tabular}{ccccc}
$n$ & HNF & SNF & superlattices & labelings \\
\hline
2  &  7   & 1   &  3  & 3 \\
3 &  13   & 1  &   3  & 3 \\
4 &  35   & 2  &   9  & 15 \\
\hline
total & 55 & 4 & 15 & 21 \\
\end{tabular}
\caption{Simple cubic superstructures for $n\leq4$.\label{tab:sc}}
\end{table}

\begin{figure*}[!t]
\noindent
\parbox{0.25\linewidth}{
  \caption{The first 21 binary structures derived from the simple
    cubic lattice. All have 4 atoms/cell or less.  Structures marked
    with a crystallographic direction $(hkl)$ can be characterized as
    a stacking of pure A and B atomic layers (black and white
    spheres. For $n=4$ all of the stacked structures occur in pairs,
    A$_3$B$_1$ and A$_2$B$_2$. The last three structures, labeled 19,
    20, and 21, cannot be characterized as pure stackings. These structures
    have basis vectors whose corresponding quotient group is
    $G=\mathbf{Z}_2\oplus\mathbf{Z}_2$ (rather than $\mathbf{Z}_4$). \label{fig:21}}}\hfill
\parbox{.73\linewidth}{\includegraphics[width=\linewidth]{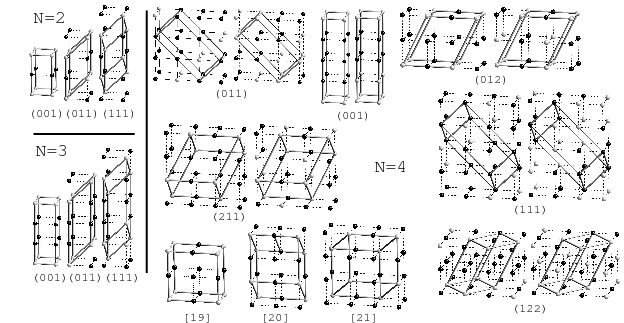} 
}
\end{figure*}

Table~\ref{tab:sc} lists the number of superlattices and superstructures for the
simple cubic lattice when $n\leq4$. The corresponding structures are visualized
in Fig.~\ref{fig:21} (compare this to Fig.~\ref{fig:17}). There are more simple
cubic derivitive structures than fcc/bcc because there are more
superlattices for a simple cubic parent lattice than for a fcc/bcc parent. 

Similar to the fcc case shown in Fig~\ref{fig:17}, most of the simple
cubic superstructures can be characterized as stackings of pure A and
B planes. The stacking directions are indicated in the figure. In
contrast to the fcc case, there are 6 unique stacking directions. It
is interesting to note that the three structures that cannot be
characterized as pure stackings are the only ones corresponding to a
composite quotient group, namely
$G=\mathbf{Z}_2\oplus\mathbf{Z}_2$. This is also true for the
non-stacked structures in the fcc case (Fig.~\ref{fig:17}), L1$_2$ and
AgPd$_3$. For the `stackable' structures, the quotient group is a single
cyclic group, $\mathbf{Z}_4$.

Table~\ref{tab:tern_quat} lists the number of fcc/bcc \emph{ternary}
and \emph{quaternary} derivative structures. A figure displaying the
ternary structures for $n\leq4$ is unnecessary---the ternary
structures have the same unit cell as the binary structures,
only the labelings are different. For $n=3$ the labeling $aab$ is
replaced by $abc$. For $n=4$ the labelings $aaab$ and $aabb$ are
replaced by $aabc$ and $abac$. And, the Agd$_3$ structure has both
labelings, rather than one. 

\begin{table}
\begin{tabular}{r|rr|rr}
$n$&\multicolumn{2}{c}{ternary}&\multicolumn{2}{c}{quaternary}\\
&structures&cumulative&structures&cumulative\\
\hline
   3   &        3   &        3 &       --  &       --\\
   4   &       13   &       16 &       7   &        7\\
   5   &       23   &       39 &       9   &       16\\
   6   &      130   &      169 &     110   &      126\\
   7   &      197   &      366 &     211   &      337\\
   8   &     1\,267 &     1\,633 &  2\,110 &     2\,447\\
   9   &     2\,322 &     3\,955 &  5\,471 &     7\,918\\
  10   &     9\,332 &    13\,287 & 32\,362 &    40\,280\\
\end{tabular}
\caption{ Number of ternary and quaternary derivative structures for an fcc parent
  lattice. Compare to the number of binary structures of Table~\ref{tab:tot_enum}.
 As the number of labels $k$ is increased, the number of derivative structures 
increases rapidly. \label{tab:tern_quat}}
\end{table}
\section{Applications}
\label{sec:apps}
We give here a few examples of how the list of derivative
superstructures can be used. An in-depth discussion of each example
is beyond the scope of this paper. The full results of each example
will be reported in forthcoming publications but we summarize our
findings here to demonstrate how the efficient enumeration of
derivative structures can be applied to discover new physics or aid in
materials design.

\subsection{fcc example}
Figure~\ref{fig:17} shows small-unit-cell binary superstructures for an fcc
parent lattice enumerated with our algorithm. There are 17 unique structures in
the set, two for $n=2$, three for $n=3$, and 12 for $n=4$. Most of the
structures can be envisioned as stackings, along different crystallographic
directions, of layers containing a single kind of atom. For instance, the $n=2$
structure labeled L1$_0$ can be envisaged as A and B layers alternately stacked
in the [001] direction. The stacking directions are indicated by green planes in
the figure.

Of these 17 small-unit-cell fcc superstructures, some are well-known structures
and others have never been observed (blue shading). Some have never been
observed experimentally but are predicted\cite{stefano_monster} to be
thermodynamically stable at low temperatures (yellow shading). It is intriguing
to ask why, among these small, simple structures, some are observed in Nature
and others are not. Recently, a metric of `bonding randomness' was
proposed\cite{missing} that shows structures with low bonding randomness are
more likely to be observed in Nature.

In the case of these 17 fcc-derived superstructures, those that have been
observed (unshaded) are generally `unrandom', those that that have never been
observed or even predicted to exist (blue shading) have a highly random bonding,
and those in between are those that have not yet been observed but have been
predicted to exist (see Fig.~4 in Ref.~\onlinecite{missing}). Noteworthy among
these results is the case of the structure predicted to exist in Cd-Pt at a 1:3
stoichiometry (see Fig.~\ref{fig:17}). We refer to this structure as L1$_3$ in
analogy to the \emph{Strukturbericht} designations for related structures.

The first theoretical discovery of the L1$_3$ structure was by M\"{u}ller in
Ag-Pd.\cite{stefan_L13} Later, a datamining search\cite{stefano_monster} found
L1$_3$ as a candidate ground state in two binary systems, Cd-Pt and
Pd-Pt. Subsequently, we constructed cluster expansions for these two systems so
that the energy of any configuration could be computed quickly. Then, by
enumerating all possible configurations, we performed (practically) exhaustive
searches of fcc-derived superstructures and verified that this structure is
globally stable in these two systems.

Because the L1$_3$ structure had not been observed and was not even suspected,
the computational discovery of L1$_3$ could have never occurred without
exhaustively examining derivative superstructures as we have done here. Without
the enumeration, this structure never would have been considered in the
searches. Our new algorithm provides a faster, more general way to enumerate
derivative superstructures. It has turned up a number of unsuspected structures with low
bonding randomness, which are good candidates for new structures that are likely
to be observed experimentally.

\begin{figure}
\includegraphics[width=.5\linewidth]{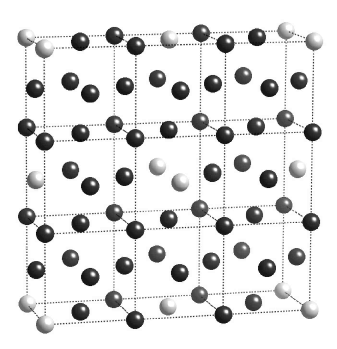}
  \caption{The fcc-derived superstructure of Pt$_8$Ti. The structures
    is shown as a face-centered tetragonal structure. The dotted lines
    show the conventional fcc cell. Black spheres represent Pt, white
    spheres Ti. This structure has the lowest bonding randomness of all 9
    atom/cell fcc superstructures.
  \label{fig:pt8ti}}
\end{figure}

\subsection{8:1 stoichiometry example}
In this example we apply the enumeration algorithm to structures with a fixed
concentration.  A structure with an unusual stoichiometry of 8:1 was
discovered\cite{pt8ti} in Pt-Ti in the late 1950s (see
Fig.~\ref{fig:pt8ti}). The structure was then `re-discovered'\cite{ni8nb,ni8nb2}
in Ni-Nb in 1969, and has since been observed in about a dozen more
Pt/Pd/Ni-rich intermetallic systems. This structure has practical import because
its impact on the alloy. Pure platinum and palladium are quite soft but the
Pt$_8$Ti structure can significantly strengthen the material, while at the same
time maintaining the purity of the host material. (International jewelery
hallmarking standards require that the alloy be at least 95 wt-\% pure.)

It would be useful to know which Pt/Pd-rich compounds can form stable
structures and what the structures are. Using the structural
enumeration method described in this paper, we generated all
9-atom/cell, 8:1 stoichiometry structures---there are 14 such
structures.  Taking these enumerated structures, we
ranked\cite{missing} the structures and find that the Pt$_8$Ti
structure has the highest likelihood among the 14. However, one of the
other 13 structures has a likelihood index nearly as high as
Pt$_8$Ti. Therefore it is another candidate for a new Pt/Pd/Ni-rich
structures. Using first-principles methods, combined with the
enumeration algorithm, we are conducting a broad search for the other
high-likelihood Pt/Pd-rich structures. So far we have found 9 new
binary systems that have undiscovered Pt- or Pd-rich ground
states.\cite{erin_aps} 

\subsection{Perovskite example}

Because our enumeration algorithm works for any parent lattice (i.e., it is 
independent of geometery), we can apply it to cases outside the realm of normal
alloy questions (usually fcc- or bcc-based systems). This example demonstrates
its use in ordering problems where the parent lattice is simple cubic. There is
a large class of oxide compounds where the `configuration question' is based on
a simple cubic parent lattice.

These oxide compounds have a number of important properties such as
ferroelectricity and superconductivity. Because of their important scientific
and industrial applications, there is motivation to optimize their functional
properties to meet different design needs. The first approach to optimizing
their properties is to change the constituent elements. This approach is limited
by the number of elements that will form the perovskite structure so to further
tune the properties, researchers create mixtures of two (or more) different
perovskites.

These oxides have the perovskites structure (see Fig.~\ref{perovskite} ). The
formula unit for this structure is ABO$_3$ where A and B are typically
transition metals or late-period metals like Pb. The A atoms are
octahedrally-coordinated by oxygen atoms, and the A-sites form a simple cubic
lattice. The B-sites also form a simple cubic lattice in the interstitial sites
of the oxygen octahedra.

\begin{figure}
\includegraphics[width=.5\linewidth]{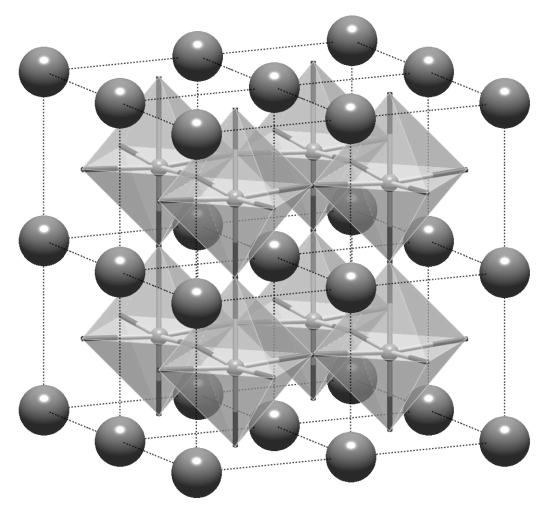}
  \caption{Depiction of the perovskite structures. B-sites are the cube corners,
    A-sites the cube centers. The oxygen atoms (not shown) form an octahedron
    around the A-sites.
  \label{fig:perovskite}}
\end{figure}

In an attempt to tune the properties of perovskite materials, researchers often
introduce a third metal. For B-site mixing, the formula unit then becomes
AB$_{1-x}$B$'_x$O$_3$. The atoms on the substitution sites will often order,
resulting in a superstructure larger than the original cubic unit cell. Because
the ordering also affects the materials properties (sometimes drastically), it
is helpful to know what ordered structures are possible and which are
likely. This is precisely the information that our algorithm gives.

The B-sites of the perovskite structure form a simple cubic lattice as shown in
the figure (big spheres). To find all possible orderings of B and B$'$ atoms on
the B site, we apply our algorithm to the case of a simple cubic
superstructures. For unit cells with 4 B-atoms or less per unit cell, we find 21
possible structures (see Table~\ref{tab:sc} and
Fig.~\ref{fig:21}).
%

A number of these superstructures have been
observed.\cite{perov1,perov2,perov3,perov4,perov5} For $n=2$, all three types
have been observed. For $n=3$, the (111) oriented structure has been
observed. The only $n=4$ structure that we are aware of is the A$_3$B$_1$
structure that we have labeled [21] in Fig.~\ref{fig:21}. Apparently, only these five
structures have been observed in B-site mixed perovskites. Whether or not there
are other structures that are likely awaits further exploration, but our
algorithm provides a starting point for finding complex perovskites with new
structures.

\subsection{Ternary fcc structures}

The two important generalizations of our algorithm over the FWZ method are
arbitrary parent lattices and arbitrary numbers of labels (atom types). The
second means we can treat cases beyond binaries, $k>2$. In the following example
we exploit the latter generalization to examine ordered ternary intermetallics.

Ordered binary intermetallics offer advantages over conventional disordered
alloys for structural applications at high temperatures. But they also have
disadvantages. Although they typically have high moduli, melting points, and
strain hardening rates, they suffer from low ductility and brittle failure. One
would like to develop alloys which can retain the advantages but avoid the
problem of poor ductility.

Cubic L1$_2$-like alloys often meet the ductility requirement but unfortunately
occur in nickel, cobalt, and platinum based binary alloys which are too dense
for the desired applications. Aluminum-rich alloys would therefore be attractive
but for the fact that most binary aluminum compounds are not cubic. Thus we need
a way to form cubic alloys with lightweight constituents.

One promising avenue is multicomponent alloys ($k>2$), primarily
ternaries. Adding a third component to a binary alloy that forms L1$_2$-like
structures such as D0$_{22}$ often induces a cubic phase that is more
ductile. Although much work has pursued this avenue,\cite{note_ternary,kumar}
there are significant experimental difficulties in identifying phases,
determining sublattice compositions, and discerning crystal structure of ordered
phases. Computational modeling is complementary to the experimental effort.

Our algorithm/code provides an important component to modeling efforts. Coupled
with a ternary cluster expansion\cite{cwolverton:1994,gceder:1994} code we
recently developed, we can now exhaustively explore the space of possible
structures in ternary intermetallics. The number of ternary fcc superstructures
is vastly larger than the number of binary structures (compare
Table~\ref{tab:tern_quat} and Table~\ref{tab:tot_enum}), so a systematic
approach like ours is even more important. The enumerated ternary structures
would also be useful in a datamining approach like that of
Ref.~\onlinecite{stefano_monster}.


\section{Summary}
We developed an algorithm for enumerating derivative structures. The algorithm
first generates all unique superlattices by enumerating all Hermite normal form
matrices and using the symmetry operations of the parent lattice to eliminate
rotationally equivalent superlattices. Next, the algorithm generates all
possible atomic configurations (labelings) of each superlattice and eliminates
duplicates using a group-theoretical approach rather than geometric analysis.

The algorithm is exceptionally efficient due to the use of (i) perfect, minimal
hash tables and (ii) a group-theoretical approach to eliminating duplicate
structures. These two features result in a linearly scaling algorithm that is
orders of magnitude faster than the previous method. Moreover, the method can be
applied to any parent lattice and to arbitrary $k$-nary systems (binary,
ternary, quaternary, etc.). The method is formally complete (does not
undercount) and key parts of the algorithm (and its implementation) can be
rigorously checked by number theory results and Burnside's lemma.

We presented results for the number of superlattices and derivative structures
for several different parent lattices as well as binary, ternary and quaternary
systems. Additionally, applications of the algorithm to alloys were
demonstrated. We coupled the results with cluster expansions and the geometric
approach of Ref.~\cite{missing} and (i) provided evidence of an altogether new
intermetallic structure, L1$_3$, in Cd-Pt and Pd-Pt, (ii) found the novel
Pt$_8$Ti to be the most likely structure for 8:1 fcc derivative structures and
uncovered another possible new prototype, (iii) enumerated small-unit-cell
structures of perovskite alloys, providing possible new structures for materials
design, (iv) demonstrated how the method can be applied to aid ternary alloy
design.

\section{Acknowledgments} G. L. W. H. gratefully acknowledges financial support
from the National Science Foundation through Grant
No. DMR-0650406. G. L. W. H. thanks Martin L. Searcy and Bronson S.  Argyle for
useful discussions concerning algorithm development and implementation of hash
tables and hash functions. We also wish to thank Axel van de Walle whose input
was helpful in testing the code.

\section{Author Contributions} G. L. W. H. conceived the project, implemented
the algorithm in Fortran 90, and wrote the first draft of the paper. The
algorithm was developed jointly but key components were invented by R. W. F,
who also implemented the algorithm in Maple as a check on the Fortran
implementation. Both authors contributed to the manuscript.
\section{Appendix}

\emph{Hermite Normal Form:} If $L$ is a lattice, with basis given by the columns of a square matrix
$\bb A$, 
and $S$ is a superlattice, then $S$ will have basis $\bb{AM}$ where $\bb M$ is
a square matrix of integers.  Furthermore, all bases of $S$ will have the
form $\bb{AMN}$ where $\bb N$ is an integer matrix with determinant $\pm 1$.
Thus, to find a canonical basis for $S$, we may use elementary
{\it integer} column operations on $\bb M$ to make it lower triangular, with
positive entries down the diagonal.  Furthermore,
we can arrange that the lower triangular matrix $\bb H=\bb{MN}$ have the property
that every off-diagonal element is less than the diagonal element in its row.
Such a matrix $\bb H$ is said to be in Hermite Normal Form, and is unique
with respect to the matrix $\bb M$.  

Thus, if the determinant of $\bb M$ is $n$, then the number of superlattices $S$
of $L$ with index $n$ is equal to the number of distinct HNF matrices with
determinant $n$.  In 3 dimensions, that number is
$$\sum_{d\big\vert n}d\>\sigma(d)=
\prod_{i=1}^k\Biggl({(p_i^{e_i+2}-1)(p_i^{e_i+1}-1)\over(p_i-1)^2(p_i+1)}\Biggr),$$
where $n=\prod p_i^{e_i}$ is the prime factorization of $n$.\smallskip

\emph{Smith Normal Form:} Using elementary integer row and column
operations (adding or subtracting an integer multiple of one row or column to
another, multiplying a row or column by $\pm 1$, or exchanging two rows or
columns), we may reduce the integer matrix $\bb M$ to a diagonal matrix $\bb D$ with the
following properties:

(i) Each diagonal entry of $\bb D$ divides the next one down.

(ii) The product of the diagonal entries of $\bb D$ is the absolute value of the
determinant of $\bb M$.

\noindent
This is called the {\it Smith Normal Form} of $\bb M$.  In the lattice case,
where $L$ is a lattice with basis $\bb A$, and $S$ is a superlattice (subgroup)
with basis $\bb{AM}$, then $\bb D$ describes the quotient group $L/S$ as
a direct sum of cyclic groups.  The
diagonal entries of $\bb D$ are the orders of the cyclic direct summands
of the quotient group (as in the Fundamental Theorem of Finite Abelian
Groups).  For example, using the notation ${\bf Z}_n={\bf Z}/n{\bf Z}$,
if $D=\begin{pmatrix}D_{11}&0&0\cr 0&D_{22}&0\cr 0&0&D_{33}\cr\end{pmatrix}$, then
$$L/S\cong G={\bf Z}_{D_{11}}\oplus {\bf Z}_{D_{22}}\oplus {\bf Z}_{D_{33}}.$$

A simple, two dimensional example, may help to show how this affects
our lattice labeling problem.  Consider the three matrices (all in HNF form)
$$\bb H_0=\begin{pmatrix}2&0\cr 0&4\cr\end{pmatrix}\hskip .3in \bb
H_1=\begin{pmatrix}2&0\cr 1&4\cr \end{pmatrix}\hskip .3in \bb
H_2=\begin{pmatrix}2&0\cr 2&4\cr\end{pmatrix}.$$ The matrices $\bb H_0$ and $\bb
H_2$ both reduce to the SNF matrix $\begin{pmatrix}2&0\cr 0&4\cr\end{pmatrix}$,
which corresponds to a quotient group which is Abelian, but not cyclic, but the
middle matrix, $\bb H_1$, reduces to SNF matrix $\begin{pmatrix}1&0\cr
  0&8\cr\end{pmatrix}$, corresponding to the cyclic group of order 8.  Thus, if
we take $\bb A$ to be the identity matrix, so $L={\bf Z}^2$, and let $S_i$ be
the lattice with basis $\bb H_i$, then $L/S_0$ and $L/S_2$ are each isomorphic
to the group ${\bf Z}_2\oplus {\bf Z}_4$, while $L/S_1$ is isomorphic to the
cyclic group of order 8.

\begin{figure}
\includegraphics[width=\linewidth]{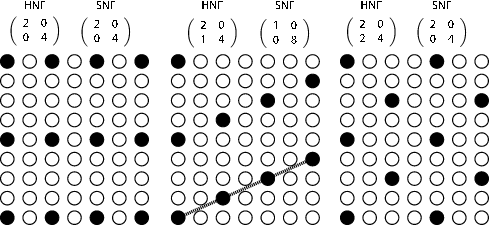}
  \caption{Two-dimensional superlattices of index $n=8$. The first and third
    share the same SNF while the second has the trivial SNF, implying a purely
    cyclic quotient group. In contrast to the first and third superlattices,
    the lattice points in layers parallel to the superlattice edge (dotted line)
    must all
    have the same label.   
    \label{fig:2D}}
\end{figure}

The fact that the latter is cyclic means we can {\it layer} the parent lattice
in such a way that each parallel layer consists of points which all must get the
same label (see Fig.~\ref{fig:2D}). We can arbitrarily label each layer passing
through the interior points of the basis parallelogram, and label the rest of
the lattice cyclically, as if labeling a one-dimensional lattice. The quotient
group for $S_0$ or $S_2$ is not cyclic but can just as easily be used to
determine equivalent labelings.

In general, SNF provides a natural homomorphism from the parent lattice $L$ onto
the direct-sum group $G={\bf Z}_{D_{11}}\oplus {\bf Z}_{D_{22}}\oplus {\bf
  Z}_{D_{33}}$, with kernel $S$. By the First Homomorphism Theorem, it
effectively gives an isomorphism from $L/S$ to $G$.  Since we do only elementary
integer row and column operations, we may write $\bb D=\bb{PMQ}$, where the transition
matrices $\bb P$ and $\bb Q$ are integer matrices with determinant $\pm 1$.  Note that
$\bb{AMQ}$ is another basis for $S$, so an element $x\in S$ iff $\bb{MQ}z=\bb A^{-1}x$ for
some integer column vector $z$, which is true iff $\bb Dz=\bb{PA}^{-1}x$.  So the map
$$x\;\mapsto \; \bb{PA}^{-1}x\;\Biggl({\rm mod}\; \begin{pmatrix}D_{11}\cr D_{22}\cr D_{33}\cr\end{pmatrix}\Biggr),$$
(meaning that each row of the resulting column matrix is reduced modulo the corresponding diagonal
element of $\bb D$) maps from $L$ into the direct sum group
$G$, with its kernel being the superlattice $S$.  

As for computing the SNF of a matrix, there are special algorithms designed to
compute it efficiently when $\bb M$ is very large but the simplest algorithm,
effective for small (e.g., 3$\times$3) matrices, is basically an extension of
Euclid's algorithm for finding the greatest common divisor of two numbers.  One
subtracts multiples of elements in the matrix from other elements in the same
row or column (using column or row operations respectively) until the greatest
common divisor of all the elements of $\bb M$ is exposed.  That element is then
moved to the upper left corner of the matrix and used to zero out all other
elements in the first row and in the first column.  Then one applies the same
algorithm to the 2 by 2 submatrix in the lower right.  Thus, in particular, the
upper left entry in $\bb D$ is always the greatest common divisor of all the
entries in $\bb M$.

Note that the number of $3\times3$ SNF matrices with determinant $n$ is
given by $\prod_i P_3(e_i)$, where $n=\prod_i p_i^{e_i}$ (the prime factorization) and
$P_3(k)$ is the number of partitions of an integer $k$ using {\it at most
3 summands}.\cite{note_partition}


\begin{thebibliography}{41}
\bibitem[{not({\natexlab{b}})}]{note3}
\bibinfo{note}{We believe that these restrictions in the FWZ algorithm have
  been overcome in the other implementations cited above, though there are no
  publications that discuss these modern implementations explicitly. However,
  our algorithm is conceptually different and completely general---geometrical
  considerations are not central to the enumeration as in the FWZ approach.
  Moreover, our approach is far more efficient and scales linearly---the
  running time of the algorithm is directly propotional to the the number of
  unique structures found.}
\bibitem[{not({\natexlab{c}})}]{note4}
\bibinfo{note}{In the implementations of
  Refs.~\onlinecite{avdwalle:02,avdwalle:02b,zarkevich:104203}, the structures
  appear to be compared directly in a geometric sense, requiring an $O(N^2)$
  search. In the FWZ algorithm, a subset of the correlations between sites in
  each structure are computed and then these correlations are compared, still
  an $O(N^2)$ search. Because the subset of correlations is limited by its
  nature, this approach is formally incomplete.}
\bibitem[{ima()}]{imagegroup}
\bibinfo{note}{In three dimensions, each member of the quotient group formally
  has three components, even though in many cases only one is non-trivial. For
  example, if the diagonal elements of the SNF of a superlattice are (1, 1, 4),
  then the 4 members of the image group are (0,0,0), (0,0,1), (0,0,2), and
  (0,0,3). On the other hand, if the diagonal elements of the SNF are (1,2,2),
  then the 4 members of the image group are (0,0,0), (0,0,1), (0,1,0), and
  (0,1,1).}
\bibitem[{not({\natexlab{e}})}]{note_ternary}
\bibinfo{note}{The literature for ternary substitutions in L1$_2$-like binary
  compounds is vast. For a representative survey, see the review by
  Kumar.\cite{kumar}}.
\end{thebibliography}
\end{document}